\documentclass[aps,prl,longbibliography,twocolumn,superscriptaddress,floatfix,10pt]{revtex4-2}

\usepackage{amsmath}
\usepackage{graphicx}
\usepackage{mathtools}
\usepackage{xcolor}
\usepackage{float}
\usepackage{bm}
\usepackage[normalem]{ulem}
\usepackage{dcolumn}

\begin{document}

\title{Field-induced spin dynamics in \textit{i}-MAX Tb compound}

\newcommand{\BGU}{Department of Physics, Ben-Gurion University of the Negev, Be'er-Sheva 84105, Israel}
\newcommand{\NRCN}{Physics Department, NRCN, P.O. Box 9001, Be'er Sheva, 84190, Israel}

\newcommand{\NHMFL}{National High Magnetic Field Laboratory, Tallahassee, Florida 32310, USA}

\newcommand{\Linkoping}{Thin Film Physics Division, Department of Physics, Chemistry, and Biology (IFM), Linköping University, Linköping, Sweden}

\author{Dror Yahav}
\affiliation{\BGU}

\author{Daniel Potashnikov}
\affiliation{\NRCN}

\author{Asaf Pesach}
\affiliation{\NRCN}

\author{El'ad N. Caspi}
\affiliation{\NRCN}

\author{Quanzheng Tao}
\affiliation{\Linkoping}

\author{Johanna Rosén}
\affiliation{\Linkoping}

\author{Moshe Schechter}
\affiliation{\BGU}

\author{Ariel Maniv}
\affiliation{\NRCN}

\author{Eran Maniv}
\affiliation{\BGU}

\date{\today}

\begin{abstract}

We report a comprehensive study of spin dynamics in the (Mo$_{2/3}$Tb$_{1/3}$)$_2$AlC i-MAX compound using ac susceptibility measurements across a range of magnetic fields. Unique behaviors were observed, including spin dynamics in the kHz range between $\mu_0H\approx0.2 T - 6 T$, indicating a non-trivial superparamagnetic state, suggesting that the compound acts as a transitional system within the i-MAX family, bridging stable spin-dynamic materials and fluctuation-dominated ones. Field- and frequency-dependent magnetic phase transitions, coupled with relaxation behaviors, reveal complex interactions between spin density waves and superparamagnetic components. These findings, corroborated by $\mu$SR studies, deepen our understanding of magnetic phase diagrams and field-induced phenomena in i-MAX systems, laying the groundwork for further exploration of their unique properties and applications.

\end{abstract}

\maketitle

\section{INTRODUCTION}

\textit{i}-MAX compounds are in-plane ordered nanolaminated materials, based on the well known MAX phases \cite{Barsoum2000,Dahlkvist2017}. In MAX phases M is an early transition metals, A is an A-group element, and X can be C,N,B, and/or P having the general chemical formula M$_{n+1}$AX$_{n}$ (n=1,2,3,...). Most MAX phases crystallize in the $P6_3/mmc$ hexagonal structure, with same element layers stacked along the $c$ axis \cite{Dahlqvist2023}. These compounds have recently gained much interest, as they are base compounds for 2D MXenes samples, which might be potentially interesting from several technological aspects \cite{Naguib2021}.

\textit{i}-MAX compounds have the general chemical formula (M$_{2/3}$M'$_{1/3}$)$_{2}$AX where the M, and M' elements are ordered within the M plane \cite{Dahlkvist2017}, \cite{Tao2017}. This ordering is accompanied by a reduction in crystal symmetry from hexagonal to typically monoclinic, but still maintains the general stacking of the same element-type layers along one unique axis (c). Unlike the case of MAX phase compounds, it was recently found that rare-earth (RE) elements can be introduced into the \textit{i}-MAX M' site \cite{Tao2019}. This significantly increased the chemical, crystallographic and electronic properties diversity in the MAX phase family. Specifically, the addition of the strongly localized unpaired 4f electrons of the RE elements generates fascinating new magnetic compounds \cite{Tao2019,YangA,YangB,Sun,Chen}. As noted in \cite{Thierry}, such compounds are expected to show complicated magnetic structures, as a result of the oscillating RKKY coupling of the 4f electrons of the different RE atoms through the conducting electron sea together with possible geometric frustration due to the triangular RE lattice. Indeed, extensive measurements, done recently, have showed signs for such a complicated magnetic phase diagram for these compounds \cite{Tao2019,Thierry,Tao2022,Danny}. In addition, spin dynamics was also evidenced for various \textit{i}-MAX compounds \cite{Danny}. At zero external field, Tb and heavier RE \textit{i}-MAX samples (henceforth, RE-\textit{i}) have shown combined fast ($f>$MHz) and slow ($f<$MHz) spin dynamics. However, compounds containing lighter REs displayed an intermediate ($\approx$MHz range) spin fluctuation rate probed by $\mu$SR \cite{Danny}.

Specifically, the Tb-\textit{i} compound has shown interesting magnetic behaviour. Tb-\textit{i} presents two magnetic transitions in zero field, namely at T=20.1, 28K~\cite{Tao2019}. Using magnetization and neutron diffraction measurements, the magnetic state was shown to form an incommensurate spin density wave (SDW) incorporating a single propagation wave vector at T=28K. Below T=20K the magnetic structure becomes a collinear antiferromagnet where the magnetic unit cell is a doubling of the crystallographic unit cell along the crystallographic $b$ axis, giving rise to magnetic momemts in the a-c plane. Finally, applying a magnetic field yields an additional ferromagnetic component, in addition to the incommensurate SDW, for temperatures T=22K and below \cite{Tao2022}. 

Albeit the existence of previous work, the exact nature of the magnetic properties of \textit{i}-MAX compounds at high magnetic fields is yet to be resolved. One of the important tools that were not used to date is frequency dependent ac susceptibility as a function of temperature and magnetic field. This probe is especially useful for determining the spin dynamics and magnetic phase diagram of magnetic materials at high magnetic fields, and hence helps determine the exact magnetic structure of the measured samples \cite{Topping}, \cite{Balanda}. In addition, high magnetic field measurements enable better understanding of magnetic materials, especially those incorporating complex magnetic behaviour such as \textit{i}-MAX compounds, as the field can serve as a tool for dis-entangling degenerated electronic states. 

In the following we show the results of ac susceptibility measurements of Tb-\textit{i} for various temperatures and magnetic fields. 

\section{EXPERIMENTAL}

\subsection{Sample characterization }

\begin{table}[b]
\caption{\label{tab:table2}
Possible impurities that might be present in the \textit{i}-MAX samples. "Null" not detected in the studied sample, "SC" is Superconductor and the numbers in parenthesis are the expected transition temperatures.}
\begin{ruledtabular}
\begin{tabular}{cccccc}
 &Tb-\textit{i}
 \\
\hline
Mo$_2$C\footnotemark[1] & SC (7.2K) \\
Mo$_2$Al$_3$C\footnotemark[2] & SC (9.2K) \\
TbAl$_2$  & Null \\
Tb$_2$O$_3$& Null \\
\end{tabular}
\end{ruledtabular}
\footnotetext[1]{Mo$_2$C is a superconductor with transition temperature topping at 7.2K~\cite{Morton}.}
\footnotetext[2]{Mo$_3$Al$_2$C is a superconductor with transition temperature of 9.2K \cite{karki2010structure}}
\end{table}

Tb-\textit{i} was prepared from the same batch reported in \cite{Danny}. Because impurities may be present in these samples, we noted in Table \ref{tab:table2} some of these impurities that might be a factor in our measurements. These impurities are based on table S1 in the Supplemental Material of \cite{Danny}, and also on ac susceptibility measurements reported below.

These impurities have relatively low transition temperatures (less than 10K). Thus, they are not expected to significantly impact the ac susceptibility measurements, since the magnetic transition temperatures of Tb-\textit{i} are higher than those of the impurities. 

\begin{figure*}[!htbp]
    \centering
    \includegraphics[width=0.9\linewidth]{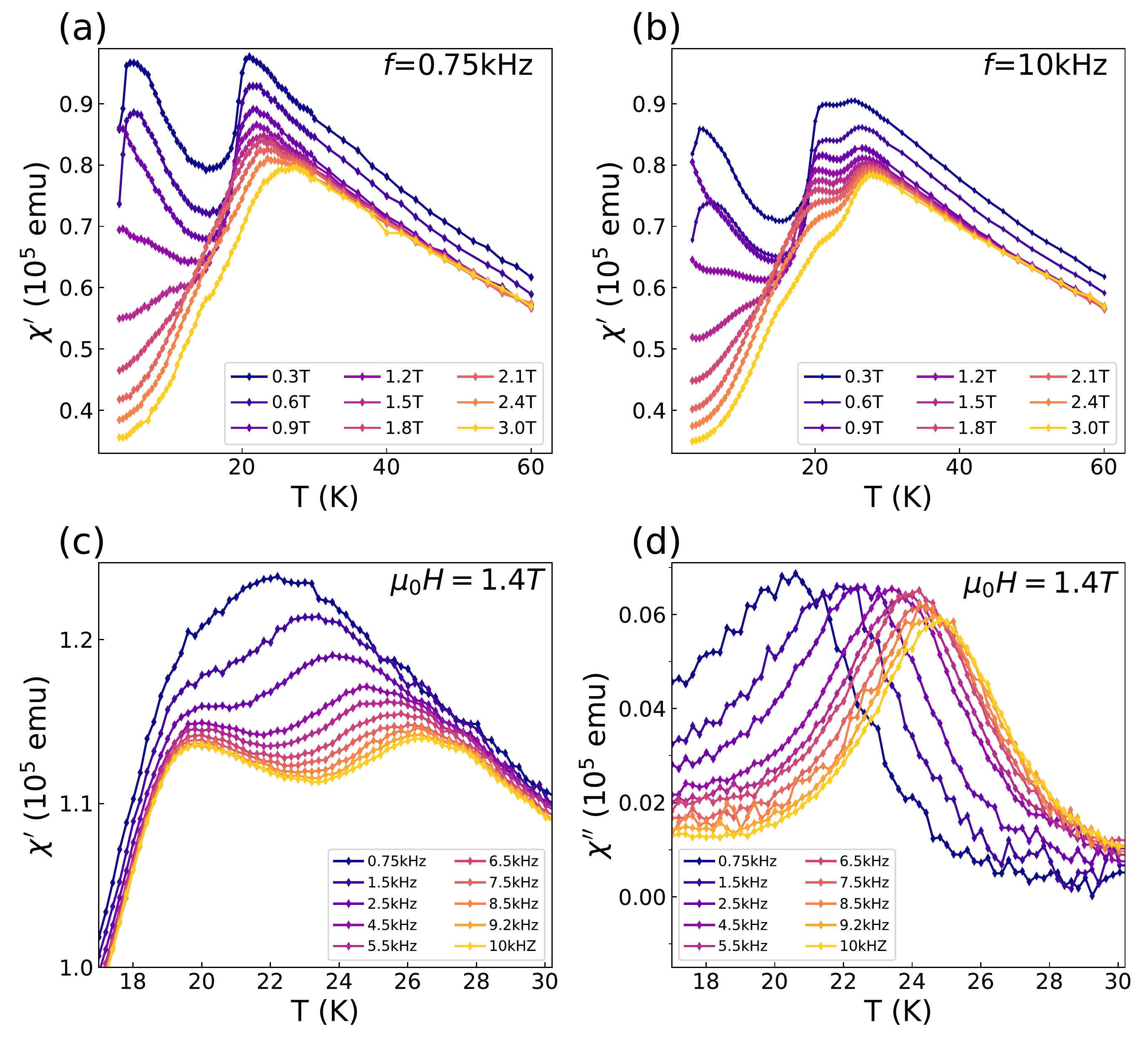}
\caption{\textbf{Evolution of the Frequency Dependent Region in Tb-\textit{i}}. Temperature scans of Tb-\textit{i} at various fields showing the evolution of the high temperature phase at frequencies of (a) $f=0.75kHz$ (low frequency) and (b) $f=10kHz$ (high frequency), displaying a single transition near $T\approx21-28K$ at low frequency versus a dual transition between $T\approx16-28K$ at high frequency. The low temperature transition becomes a kink near $\mu_0 H\approx3T$. The high temperature transition remains significant even at high magnetic fields (see Supplemental Material Figure~S2~\cite{Supp}). (c) Temperature scan for Tb-\textit{i} at $\mu_0 H=1.4T$, showing a striking difference between low and high frequencies where the low frequencies show only a single transition while high frequencies show a dual transition at $T_{c_1}\approx20K$, $T_{c_2}\approx26.5K$. The high temperature transition shows clear frequency dependence, in contrast to the low temperature constant transition. (d) The imaginary part of the susceptibility at $\mu_0 H=1.4T$, displaying a frequency dependent peak, shifting to higher temperatures while the frequency increases, characterized by a Mydosh parameter \cite{Mydosh}.}
    \label{fig:Tb low 0.75vs10kHz}
\end{figure*}

\subsection{AC susceptibility measurements}

In order to thoroughly characterize the high field magnetic phase diagram of Tb-\textit{i} sample, and track the dynamics of this compound, use was made of the ac susceptibility technique. AC susceptibility was measured using the PPMS ac Susceptibility option (Quantum Design) at the Physics Department of Ben-Gurion University, Israel. The measurements were done in the temperature range of $T=3 - 300K$ , in the field range $\mu_0 H=0 - 9 T$, and in the frequency range $f=10-10,000Hz$. In this work, peaks in the real and imaginary parts of the ac susceptibility measurements are associated with possible magnetic transitions.

\section{Results}

\begin{figure}[!htbp]
    \centering
    \includegraphics[height=0.75\linewidth]{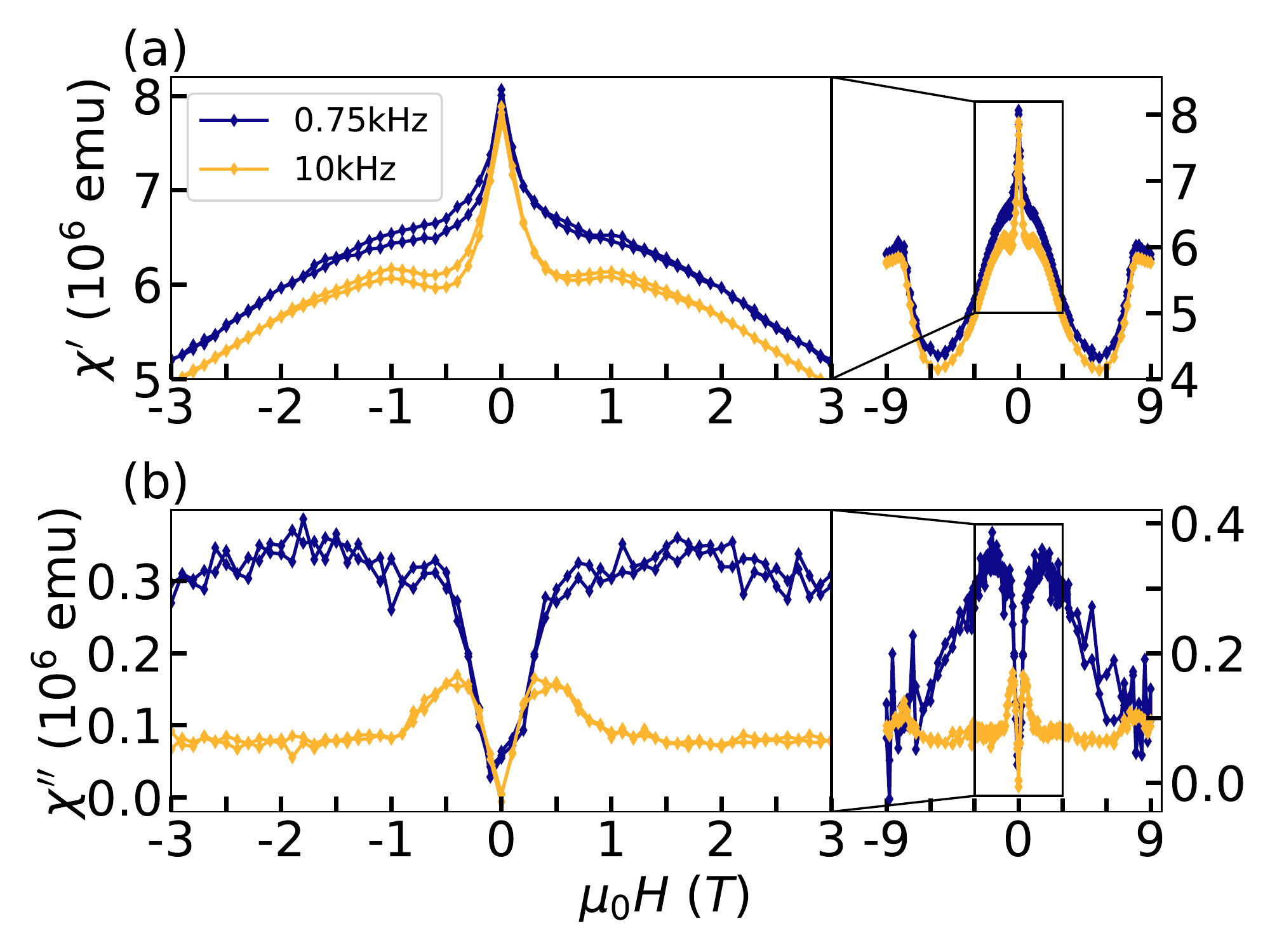}
\caption{\textbf{Field scan measurements at T=20K for both high and low frequencies}. (a) The real part of the susceptibility with a phase transition near $\mu_0 H\approx1T$ seen only for the $f=10kHz$, while for $f=0.75kHz$ shows a kink near the same field. Another transition was measured at $\mu_0 H\approx8.2T$ for both frequencies. (b) The imaginary part of the susceptibility. A peak is observed around $\mu_0 H\approx0.5T$ for  $f=10kHz$, compared to $f=0.75kHz$ were a much broader peak is seen at $\mu_0 H\approx1.8T$. The striking differences between different frequency measurements of the imaginary part lasts up to H=6T. The left panels are zoom-in of the right panels.}
\label{fig:Field scan at 20K}
\end{figure}

AC susceptibility measurements were done for the Tb-\textit{i} sample. Typical temperature dependent measurements are shown in Figure \ref{fig:Tb low 0.75vs10kHz} for several fields, focusing on the two frequencies most used during the measurements $(f=0.75kHz, 10kHz)$. At zero field a single magnetic transition has been measured between 10K to 28K (Figure~S1), initially correlated to the T=20K transition noted in the literature\cite{Tao2019,Thierry,Tao2022,Danny}. The higher (T=28K) transition noted in the literature\cite{Tao2019,Thierry,Tao2022,Danny} was not evidenced in the present (ac susceptibility) measurements at zero field. However, for the H=0.2-6T field range, a frequency dependence appears. This frequency dependence is most pronounced in the field range $0.2T<\mu_0H<3T$, in which splitting of the magnetic transition is clearly observed while increasing the frequency. This frequency dependence is also seen for the field sweep measurements shown in Figure \ref{fig:Field scan at 20K}. Note specifically the drastic change between the measurements at different frequencies of the out of phase (imaginary) part, up to H=6T, implementing a significant spin-dynamical behaviour at this region of the phase diagram. Note that the frequency dependence was not measured below 0.2T and above 6T, nor evidenced for other i-MAX samples measured. This frequency dependence marks the possible existence of a glassy state for this compound. It might also point that Tb-\textit{i} serves as a border compound separating spin-stable compounds (light RE compounds) from spin-fluctuation ones (heavier RE), in-line with previous $\mu$SR  findings\cite{Danny}.

However, when considering the relaxation measurements, focusing on the imaginary part of the magnetic signal, a different (non-spin glass) and intriguing picture appears. First, a temperature shift of the peak as a function of frequency $(\nu)$ was measured for the imaginary part of the susceptibility $\chi''$. This shift can be characterized by the Mydosh Parameter $(\phi)$~\cite{Wei}:

\begin{equation}
\label{mydosh}
    \phi=\frac{\Delta T_f}{T_f \Delta(log(\nu))}
\end{equation}

\begin{figure}[htbp]
    \centering
    \includegraphics[width=0.85\linewidth]{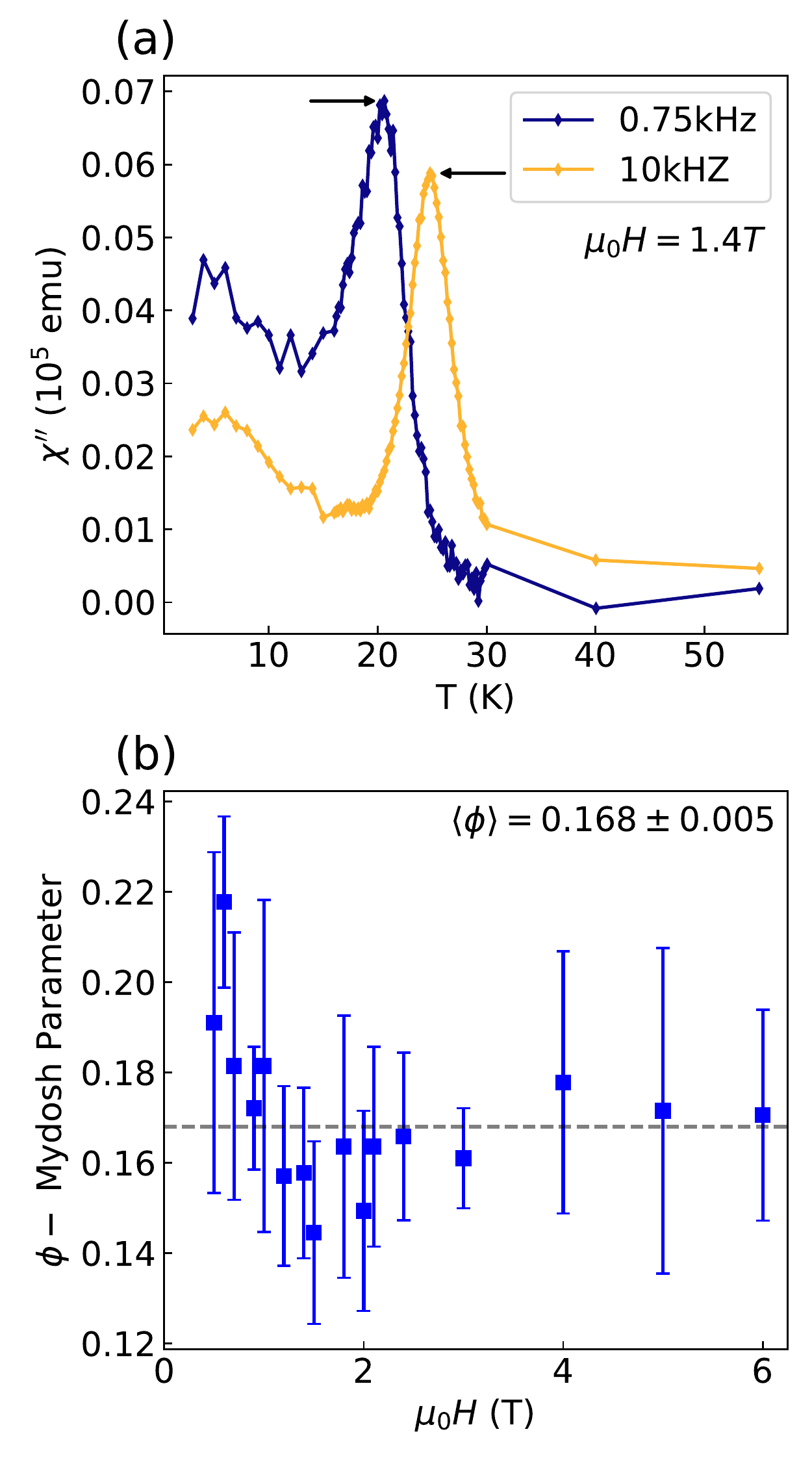}
\caption{\textbf{Mydosh Parameter}. (a) Imaginary part for $f=0.75,~10kHz$. The arrows point to the maximal loss susceptibility value of each frequency, used to calculate the Mydosh parameter calculation (see equation (1)). (b) Mydosh parameter for various fields $(\mu_0 H=0.5-6T)$ with estimated errors. Dashed line marks the average value of $\langle \phi \rangle$. Note that calculating the Mydosh parameter for fields above $\mu_0 H=7T$ is possible, but gives a large uncertainty (see Supplemental Material Figure~S4~\cite{Supp}).} 
    \label{fig:Tb 1.4t mydosh}
\end{figure}

Where $T_f$ represents the freezing temperature at which $\chi''$ reaches its maximum at the highest frequency $\nu$ (see also Figure \ref{fig:Tb 1.4t mydosh}) and $\Delta T_f$ denotes the difference in freezing temperature between the highest and lowest measured frequencies.

The Mydosh parameter $\phi$ can be considered as field-constant within the uncertainty (Figure \ref{fig:Tb 1.4t mydosh}). Note that above a field of H=6T the Mydosh parameter disappears alongside with the absence of any frequency dependence above this field (see Supplemental Material Figure~S4~\cite{Supp}). Calculating its average for the various fields measured yields: $\langle \phi \rangle=0.168(5)$. The obtained value for $\phi$ is relatively large and suggest the possible existence of a superparamagnetic state rather than a spin glass~\cite{Mydosh},~\cite{Wei}.

Afterward, the relaxation time $\tau$ was calculated by fitting the imaginary part of the susceptibility to the generalized Debye model ($\chi''(f)$):
\begin{equation}
\label{GD eq main}
    \chi ''(\omega)=\frac{\frac{1}{2}(\chi_T-\chi_S)\cos{(\pi\frac{\alpha}{2})}}{\cosh{[(1-\alpha)ln(\omega\tau)]}+\sin{(\pi\frac{\alpha}{2})}}
\end{equation}
where $\chi_T,~\chi_S$ are the isothermal and adiabatic susceptibility respectively, $\alpha$ is the parameter controlling the spread of relaxation times, $\tau$ is the characteristic relaxation time and $\omega=2\pi f$ \cite{Topping}. In the Supplemental Material~\cite{Supp} we elaborate in details regarding the relaxation model and fitting process. The fit is done for various temperatures and fields (see Supplemental Material Figure~S7~\cite{Supp} and Figure~\ref{fig:tau_temp_field}a inset), and the resultant relaxation times are presented in Figure \ref{fig:tau_temp_field}.

The temperature dependence of the relaxation time shows a double Arrhenius behaviour for $T<20K$ and $20K<T<28K$, for fields $\mu_0 H=2.0T$ and below (see Figure \ref{fig:tau_temp_field}). Note that an Arrhenius process can be described by Eq.~\ref{double_arrhenius} where $\tau_{h0}$ is the "hopping time" (or inverse attempt frequency), which can be thought of as the time between attempts at thermally exciting over the energy barrier $U_0$~\cite{Topping}.

\begin{equation}
\label{double_arrhenius}
    \tau=\tau_{h0}\cdot exp\left(\frac{U_0}{k_BT}\right)
\end{equation}
\clearpage

Specifically, at $\mu_0 H=0.2T$ a clear double Arrhenius process is evidenced, with $U_1 \approx 97K$ and $\tau_{h1} \approx 0.4\mu s$ for the high temperature process, and $U_2\approx 6K$ and $\tau_{h2}\approx60\mu s$ for the low temperature one (fitted separately). The cross over from one region to another takes place in the temperature range $15K<T<21K$ as can be seen in Figure \ref{fig:tau_temp_field}a. However, for  $\mu_0 H=0.5T$, the dynamical behaviour of $\tau(T)$ seems to be more complex with a plateau-like region (Figure \ref{fig:tau_temp_field}b). Such behavior resembles that of quantum tunneling of magnetization (QTM) (see~\cite{Hardy}), although this is highly hypothetical. Finally, increasing the field to $\mu_0 H ={1.0}T$ and above results again in an Arrhenius process, for $T\geq20K$. These drastic changes comply with the H=0.5T peak seen in field sweep measurements, evidenced only for high frequencies ($f\approx10kHz$), implying a probable magnetic transition. Table~S1 summarizes the hopping times and energy barriers extracted from temperature-frequency scans for different fields with respect to a single Arrhenius fit for high temperatures (above T=21K), in which quantitative fits were possible for all fields. The only exception is for H=0.2T, as mentioned above. 

\section{Discussion}
\begin{figure}[htbp]
    \centering
    \includegraphics[height=1\linewidth]{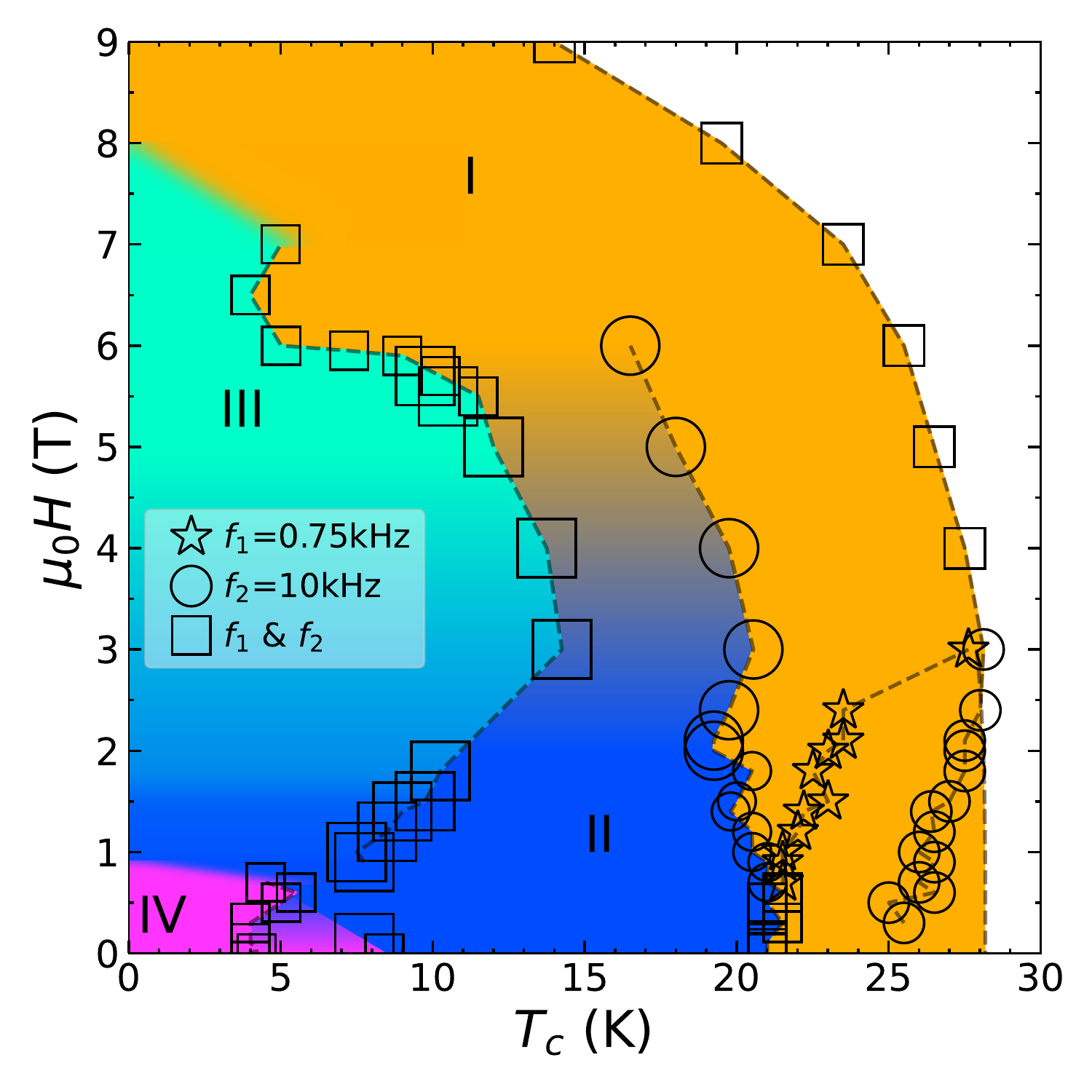}
\caption{\textbf{Phase Diagram Tb-\textit{i}}. Phase Diagram of Tb-\textit{i} compound for $f=0.75kHz$ (stars and squares) and $f=10kHz$ (circles and squares) showing 3 phases: I, II and III. Phase I is the high temperature phase, showing pronounced frequency dependence in the field range between $0.2 -3 T$. The high-frequency/ zero-temperature border of phase I is extrapolated from previous reports~\cite{Tao2019, Danny}. Phase II is the low temperature low field phase, which merges into phase I at high fields. Phase III is the low temperature phase, which also includes a superconducting parasitic phase at low fields, marked as an effective phase IV. The color gradient is a guide to the eye where the transition does not contain a local maxima and is calculated using the derivative $d\chi'(T)/dT$ rather by finding the local maximum of $\chi'(T)$. Note that the uncertainty on each measured point is included in the points size.} 
    \label{fig:PD}
\end{figure}

From the ac susceptibility measurements, a magnetic phase diagram can be constructed for the Tb-\textit{i} sample. The result is shown in Figure \ref{fig:PD}.

\begin{figure*}[htbp]
    \centering
    \includegraphics[width=\linewidth]{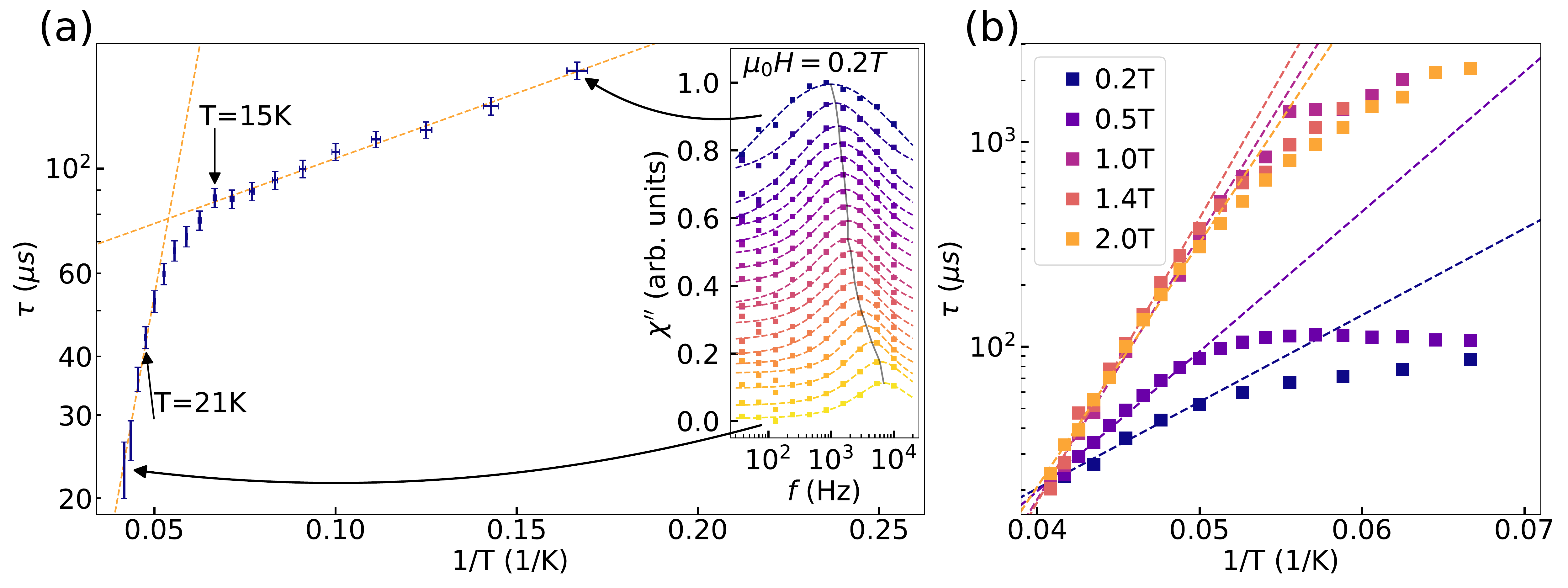}
\caption{\textbf{Relaxation Measurements.} (a) Arrhenius-like plot of $\tau=1/2\pi f$ the relaxation time as a function of the inverted temperature at $\mu_0 H=0.2T$. The relaxation time $\tau$ is calculated by fitting the generalized Debye model to $\chi''(f)$ (see Eq.~\ref{GD eq main}). Two Arrhenius fits are presented by dashed lines, showing a transition between 2 distinct relaxation processes around $T\approx 18K$. The inset shows the imaginary part of the susceptibility as a function of frequency, for different temperatures in the range of $T=6-27K$ with spacing of 1K between each measurement. Measurements were done in the range of $f=37.7Hz-10,000Hz$. The presented data is offset on the y-axis for clarity. A shift towards higher frequencies is clearly seen. From fitting Eq.~\ref{GD eq main} to $\chi''(f)$ (dashed lines), the relaxation time  $\tau$ is extracted for each temperature, and then used to draw the main Figure, as pointed by the black arrows. (b) Arrhenius-like plot of $\tau$ for various fields, showing the growth of the slope up-to $\mu_0 H=1T$ for $T\geq15K$.}
    \label{fig:tau_temp_field}
\end{figure*}

The phase diagram is separated into 3 different major phases: I, II and III. Phases I and II are the well-known magnetic phases noted in the literature ~\cite{Tao2019},~\cite{Danny}, which have been shown to be both SDW. Phase III has also been noted in the literature~\cite{Tao2022}. Specifically it complies with combined ferromagnetic (FM) and anti-ferromagnetic (AFM) SDW components of the low temperature phase extracted from neutron scattering measurements at a field of H=6T and T=5K~\cite{Tao2022}. We note that phases I and II also include a FM component, as implied by the high field neutron scattering measurements done previously~\cite{Tao2022}. The main difference between the different magnetic phases is the exact mixture of AFM/SDW and FM components, and also the exact SDW state, including either a single or double propagation wave, as noted in ~\cite{Danny}. Finally, the phase diagram includes a superconducting parasitic phase at its low field - low temperature corner, marked as an effective phase IV. This low temperature parasitic magnetic transition disappears at an external field of $\mu_0 H=0.9T$.

Starting from a field of $\mu_0 H \approx 0.2T$, a frequency dependent signal appears. This frequency dependence is evidenced via a split of phase I (single peak), which is measured in the field range between $0<\mu_0 H<0.2T$, into phases I, II, marked by a double peak structure while increasing the ac frequency (see Figure~\ref{fig:Tb low 0.75vs10kHz}). The dependence on frequency is most significant near H=1.4T, where a split in the higher peak is evidenced at a frequency of $f=10kHz$ as compared to a single peak measured for $f=0.75kHz$. This "quasi-magnetic state transition" is marked in the phase diagram by stars. Note that the spin-dynamical phase stretches for the entire region $0.2T<H<6T$ and below T=28K, at least down to T=20K. The difference between different frequency measurements, points to a significant spin-dissipation already at the kHz frequency range. As written above, the frequency dependence stretches up to a field of $\mu_0 H=6T$, above which no frequency dependence is shown. Note also that the high temperature splited phase (phase I) matches the T=28K transition measured in the literature at zero field. In addition, the low temperature phase III exist up to a field of $\mu_0 H=7T$, above which it disappears. Similarly, the higher magnetic phase I shows a gradual suppression starting from a field of $\mu_0 H=4T$.

The dynamical phase splitting of phase I  at high frequencies implies the existence of superparamgnetism at this specific region of the phase diagram, as also inferred by the value of the Mydosh parameter. In this context, curiously, the fitted attempt frequency and energy barrier increase significantly with field. This might imply increased energy level splitting induced by elevating the external field. In addition, the disappearance of the dynamic phase at $\mu_0 H=6T$ is clearly evidenced in Figure \ref{fig:PD}, where phase II seems to merge into phase I, alongside the disappearance of the Mydosh parameter. In this context, note that in \cite{Tao2019} and \cite{YangA} the hysteresis loop terminates near $\mu_0 H\approx 6T$ in the field scan for the Tb-\textit{i} sample.

The high temperature phase (phase I) is reminiscent of systems showing creation of superparamagnetic domains with typical flip time~\cite{eiselt1979magnetic}. This gives a frequency dependent response for phase I, as shown specifically in Figure ~\ref{fig:Tb low 0.75vs10kHz}c, with the corresponding Mydosh parameter (see Figure~\ref{fig:Tb 1.4t mydosh}b). As a consequence, phase II may be associated with interactions between the superparamagnetic domains, resulting in much slower dynamics as was suggested in~\cite{eiselt1979magnetic}, supporting non-frequency dependence in this region. As a result, a combined SDW with superparamagnetic domains is proposed for the spin texture of the magnetic phases at this region.

Based on the above explanation, we hypothesize that the frequency dependence actually stretches down to zero field, not measured due to the upper limit of the ac susceptometer, implying spin dynamics in the MHz region, as also measured via $\mu$SR \cite{Danny}. Such hypothesize also implies that the actual zero field transition (up to $\mu_0 H\approx 0.2T$) measured by us is related to both phase I and phase II (see supplemental Material Figure~S1~\cite{Supp}). The decrease in phase I transition temperature relative to the literature is a consequence of frequency dependence, evident in Figure~\ref{fig:Tb low 0.75vs10kHz}c and in the evolution of the relaxation time $\tau$ with increasing field (y-axis of Figure~\ref{fig:tau_temp_field}b).

\section{Conclusions}
We have measured extensively the Tb-\textit{i} compound ac susceptibility signal as a function of both temperature and field. This compound shows spin dynamics in the kHz range, implementing a probable superparamagnetic state in the field range $0.2T<\mu_0 H<6T$. The present measurements shed additional light on the slow dynamical component of the Tb-\textit{i} compound observed previously in $\mu SR$ measurements~\cite{Danny}. Together with previous work ~\cite{Tao2019}, ~\cite{Thierry}, ~\cite{Tao2022}, and our work on other RE-\textit{i} compounds~\cite{yahav2024hidden}, it all implies that Tb-\textit{i} is a corner compound separating between low RE compounds exhibiting relatively stable spin dynamics, as compared to higher RE compounds displaying a relatively fluctuation enhanced dynamics. Finally, more studies, theoretically and experimentally, especially at high fields, are required, in order to clarify the exact nature of \textit{i}-MAX compounds in the high field region.

\section{Acknowledgements}
Work by D.Y. and E.M. was performed with support from the European Research Council grant no. ERC-101117478, the Israeli Science Foundation grant no. ISF-885/23. D.Y, E. M. and A. M. acknowledges funding from the PAZY foundation grant no. 412/23. JR acknowledges funding from the Knut and Alice Wallenberg (KAW) Foundation for a Scholar Grant (2019.0433). Raw data is available in Zenodo~\cite{availability}. 

\newpage


%

\end{document}